\documentclass[eqsecnum,aps,prd,nofootinbib,]{revtex4}
\begin{document}

\def\beq{\begin{equation}}
\def\endeq{\end{equation}}
\def\bea{\begin{eqnarray}}
\def\endea{\end{eqnarray}}

\def\journaldata#1#2#3#4{{\it #1 } {\bf #2:} #3 (#4)}
\def\eprint#1{$\langle$#1\hbox{$\rangle$}}
\def\lto{\mathop
        {\hbox{${\lower3.8pt\hbox{$<$}}\atop{\raise0.2pt\hbox{$\sim$}}$}}}

\title{On the Status of Highly Entropic Objects}

\author{Donald Marolf}

\affiliation{Physics Department, UCSB, Santa Barbara, CA 93106.}

\author{Rafael D. Sorkin}
\affiliation{Physics Department, Syracuse University, Syracuse, NY 13244.}

\date{\today}

\begin{abstract}
It has been proposed that the entropy of any object must satisfy
fundamental (holographic or Bekenstein) bounds set by the object's size
and perhaps its energy.  However, most discussions of these bounds have
ignored the possibility that objects violating the putative bounds could
themselves become important components of Hawking radiation.
We show that this possibility cannot a priori be neglected in existing
derivations of the bounds.
Thus this effect could potentially invalidate these derivations; but
it might also lead to observational evidence for the bounds themselves.

\end{abstract}
\maketitle

\section{Introduction}

The laws of thermodynamics and the concepts of entropy ($S$) and
energy ($E$) express fundamental aspects of physics.  In the conventional
understanding, these quantities are related to each other and to
the size of an object only through the first law of thermodynamics
$dE = TdS - PdV$.
However, there have been intriguing suggestions
(see e.g. \cite{Bek73,tH,LS,FS,Bousso})
that more fundamental laws
(e.g., quantum gravity and/or string theory effects)
should change this picture.
In particular, the entropy of any object
might be bounded by some function of its size, typically characterized
by a length scale $R$ or an enclosing area $A$,
and perhaps its energy $E$.
Suggestions of this form
include Bekenstein's proposed bound \cite{Bek73},
\begin{equation} \label{Bek}
  S < \alpha RE/\hbar c,
\end{equation}
and the so-called holographic bound \cite{tH,LS},
\begin{equation} 
\label{holog}
  S < A c^3 / 4 \hbar G ,
\end{equation}
Here we have displayed the fundamental constants explicitly,
but below we use geometric units with $k_B =\hbar=c=G=1$.\footnote
{In many ways, the choice $8\pi G=1$ is more natural than $G=1$.  With
 this choice of units, the horizon entropy becomes $2\pi A$ and equation
 (\ref{holog}) reads $S<2\pi A$.}
The original version \cite{Bek73} of Bekenstein's bound has
$\alpha=2\pi$, while some subsequent discussions (e.g. \cite{Erice})
weaken the bound somewhat, enlarging $\alpha$ by a factor of order ten.

Arguments in favor of these bounds \cite{Bek73,tH,LS,Bek2000,Erice}
typically suggest that inserting or transforming bound-violating objects
into black holes leads to contradictions with the second law of
thermodynamics.  Many counter-arguments
have been given and the subject remains in a state of controversy.
The original argument \cite{Bek73} for
(\ref{Bek})
involved slowly lowering a ``box'' toward a black hole and then, at
some point, letting it fall freely through the horizon.
Counter-arguments appealing to a buoyant force exerted on the box by the
``thermal atmosphere'' of the black hole were given by Unruh and Wald in
\cite{UW}.
The question was
reconsidered recently in \cite{MS}
in the context of a resolution of the ``self accelerating box paradox''.
Under plausible assumptions as to the treatment of certain boundary
effects, it was shown in \cite{MS} that a box violating (\ref{Bek})
would make a notable contribution to the thermal atmosphere of the very
black hole with which it was supposed to violate the second law.
This contribution might be negligible far from the black hole, but would
become important in the region near the horizon from
which the box was to be dropped.
This opens the door to new effects which might provide loopholes
in the original argument of \cite{Bek73}.
A few such effects were discussed
in \cite{MS} and similar effects will be described below.

However, other arguments for a version of (\ref{Bek})
(with $\alpha$ somewhat greater than $2\pi$)
have been made in which one releases the object to fall into the
black hole from far away \cite{Erice}.
When applied to  such processes,
the comments of \cite{MS} suggest that
thermally produced copies of bound-violating objects
would be relevant even far away from the black hole.
In other words, despite the very
low Hawking temperature of any macroscopic black hole,
they suggest that
objects violating this version of (\ref{Bek})
would be Hawking radiated at a
significant rate by the particular black hole used in the argument.
We explicitly verify this suggestion below, noting that the situation
far from the black hole is under much better control than that studied
in \cite{MS}.

We then note that these considerations generalize to
any setting 
(e.g. those of \cite{HD} and \cite{Bek73,Erice})
in which the absorption of a
`highly entropic object'
by a black hole would,
in the absence of further entropy generation,
lead to a violation of the second law.
Thus, 
such highly entropic objects and their kin will be
important components of the
black hole's thermal atmosphere so that further processes will indeed occur.
We show that similar comments apply to
the holographic bound (\ref{holog}).
Finally, we suggest how this same
effect could lead to
{\it observational} evidence in favor of both (\ref{Bek}) and
(\ref{holog}) in certain regimes,
independently of whether, as a matter of principle,
these bounds necessarily hold in all possible hypothetical worlds.

We remark here that the majority of the thought experiments we consider
herein (as well as those considered in \cite{MS}) involve only
semiclassical processes which are quasistationary
{\it for the black hole},
that is to say processes in which the black hole may be treated
classically such that its mass changes only incrementally.
(An important exception is the gendankenexperiment
for deriving the holographic bound in \cite{LS}.)
In this context, 
a very general argument presented in \cite{chb,tpe}
establishes\footnote
{assuming that certain a priori divergent quantities can be
 handled appropriately.}
that no violation of
the GSL can occur if the matter outside the black hole is correctly
described by some quantum field theory.
{}From this point of view any
attempt to decrease the total entropy of a black hole by inserting
highly entropic objects is doomed in advance to fail, and the only
question is how this failure will work itself out in the given case.
Thus, a proof that one could decrease the entropy with the aid of a
certain type of
highly entropic object would amount to a proof that no such object could
exist in any self consistent quantum field theory (which extended to
curved spacetime). 
Conversely, if one could imagine a quantum field
theory in which such an object definitely could exist, then one would be
guaranteed that the theory would provide for some effect to protect the
GSL from violation,
when such objects were made to interact with black holes in the above
semiclassical setting.
To a certain extent, the remainder of this paper is just a more detailed
working out of this implication.

\section{Highly Entropic Objects at Equilibrium}

\label{SER}

It is well established \cite{UW2,Ray,BFZ,TJ1,TJ2,TJ3} that
the radiation surrounding a black hole of temperature $T_{BH}$ 
is thermal in the sense that,
{\it in equilibrium},
it is described by an ensemble of the form $e^{-\beta H}$,
where $\beta=1/T_{BH}$.
When a black hole radiates into empty space 
and the thermal ensemble would be dominated by weakly interacting particles,
the Hawking radiation is just the outgoing component of the radiation
described by this ensemble.

The point stressed in \cite{MS} is that, 
according to statistical mechanics, 
the probability to find a particular macrostate in a thermal ensemble 
is not $e^{-\beta E}$ but $e^{-\beta F}$,
where $F=E-T_{BH}S$ is its free energy\footnote%
{Since $e^{-F/T} = e^{-E/T} e^S$, the free energy includes the effect of
 collecting $e^S$ microstates into a single macrostate.} 
at the temperature $T_{BH}$ corresponding to the black hole.
In converting this into an emission rate, 
the only other relevant factor is
a `gray body factor' that enters the absorption cross section $\sigma$
for our object.  (The absorption and emission
rates
are related by the assumption of ``detailed balance''.  We assume that
this assumption is valid for our objects.\footnote%
{If the object can be described as a field quantum and the black hole
 metric treated as fixed, then one just has potential scattering, for
 which detailed balance can be derived in the usual manner.  More
 generally, one might appeal to some version of time reversal
 invariance, or better CPT invariance, but the status of the latter is
 not settled within quantum gravity.})

\subsection{The Bekenstein bound}
\label{Bb}

Let us now recall the setting for the argument of \cite{Erice} in favor
of (\ref{Bek}). 
One considers an object of size $R$, energy $E$, and entropy $S$ which
falls into a Schwarzschild black hole of size $R_{BH}=2\zeta{R}$ from a
distance $d{\gg}R_{BH}$.
The parameter $\zeta$ is taken to be large enough that the object
readily falls into the black hole without being torn apart.
In other words, we engineer the situation so that the black hole is, at
least classically, a perfect absorber of such objects.
It is also assumed that the Hawking radiation emitted during the infall
of our object is dominated by the familiar massless fields, in which
case it is a small enough effect so as not to significantly impede the
fall of our object.
Consideration of the second law \cite{Erice} then leads to the bound
\begin{equation}
                 S < 8\pi \nu \zeta R E,
\end{equation}
where $\nu$ is a numerical factor in the range
$\nu=1.35-1.64.$   
Here the energy $E$ has been assumed to be much smaller than the mass
$M_{BH}$ of the black hole and, up to the factor $\nu$, the above bound
is obtained by considering the entropy increment of the black hole,
$dS_{BH}=dE_{BH}/T_{BH}$.
Notice here that the black hole was assumed not to readily emit
copies of our object as part of its Hawking radiation.

Suppose, now, that a ``highly entropic object'' does exist with
$S > 8\pi \nu \zeta R E$.  The arguments of \cite{MS} suggest
that this large entropy will induce such objects
to be emitted copiously by the black hole, and it is clear that no
violation of the second law will result if the net flux of such objects
vanishes or is directed outward from the black hole.
To see whether this is indeed the case,
let us compute the free energy of our object at the black hole temperature
$T_{BH}=(4\pi R_{BH})^{-1}=(8\pi\zeta R)^{-1}$:
\begin{equation}
            F = E - \frac{S}{4 \pi R_{BH}} < E - \nu E < 0.
\label{Feqn}
\end{equation}

Now if we assume that  no objects are present, then we find
$F=E-TS=0-0=0$.
By (\ref{Feqn}), 
objects violating (\ref{Bek}) 
have significantly lower free energy than this, and are
therefore more likely to exist than not
in a state of thermal equilibrium at $T_{BH}$.
(In fact, the most likely macrostate is one that is so full of such
objects that new ones cannot be squeezed in at the same low free
energy.)
Consequently, it is unjustified to assume that
such objects are unlikely to be radiated by the black hole, during the
course of one of the putatively entropy-violating processes under
consideration.  (On the other hand, we cannot simply assert that they
must be radiated in great numbers, because the nature of the equilibrium
state does not in itself determine what happens away from equilibrium.)

Consideration of simple models elucidates the ways
in which this loophole might play itself out.
Suppose for example that our object's free energy were independent of
the number of such objects already present\footnote
{This supposition is instructive but not realistic.
 Note in particular that free bosons do {\it not} fall into this
 category, as a thermal ensemble of any number $N$ of free boson fields
 exists any temperature $T$.  The free boson case is quite interesting
 and will be studied in detail in \cite{MMR}}.
Then the putative thermal ensemble would be unstable, as adding an
additional such object would lower the free energy, no matter how many
were already present.
(Hence,
strictly speaking, there could be no state of equilibrium at all,
much as with the super-radiant modes in the case of a rotating black
hole.)
Thus, we would expect the Hawking radiation to contain so many of our
objects that the usual semi-classical approximation would fail and the
black hole would quickly decay.

As a second example, suppose that our objects can be modeled by
weakly interacting Fermions.  
Then all macrostates
with $F<0$
will be occupied, although states with sufficiently high kinetic energy
will remain empty.  If the parameters are chosen correctly, the rate of
Hawking radiation can remain low enough that the semiclassical
approximation remains valid and the black hole does exist as a
metastable state.  However, since the object we wish to drop is by
construction in a state
with insufficient kinetic energy to satisfy $S < 8\pi \nu \zeta R E$, it
represents an ingoing state with
$F<0$.  
Thus, 
a
corresponding outgoing state is occupied with high
probability 
and the black hole will very likely emit
such an object during the time that our ingoing object is being
absorbed.  In fact, it is very likely to emit a large number of
such objects in various directions.

In the third instructive case we suppose that the thermal atmosphere of
the black hole blocks the passage of our highly-entropic objects so that
energetic objects cannot stream freely outward from the black hole.  Let
us assume it also blocks the passage of the CPT conjugate objects, since
these will carry equal entropy.  This case might arise because the
thermal atmosphere already contains many densely packed copies of our
object, or it might arise because our objects are excluded by
interactions with some other
component of the atmosphere.
Note that due to 
detailed balance (or CPT invariance)
this
atmosphere will also obstruct us from dropping in a new object from far
away.  
Our new object will bounce off the thermal atmosphere or be
otherwise prevented from entering the black hole to the same extent that
an outgoing such object emitted by the black hole will fail to escape.
Thus, again it is plausible that the black hole is very likely to emit at
least
one such object before we manage to send a new one into the black hole.

In each case we find, with high plausibility, that the Hawking
radiation adds at least as much entropy to the universe as is removed when
our 
object falls through the horizon.
Note that none of the caveats
from \cite{MS} apply here: the relevant region is far from the black
hole so that it is large and homogeneous and no boundary effects
should 
be important.

\subsection{A generalization}

Since the end result did not rely on particular properties of
Schwarzschild black holes, one might
expect that our argument can be formulated much more generally.
To
see that this is the case,
let us proceed along the lines of \cite{chb}.
Consider then any process in which a given object with entropy $S$
is destroyed, giving its energy $E$ to a
black hole.  
Note that this includes both processes of the original form \cite{Bek73}
as well as the more recent \cite{Erice}.
As above, we suppose that this represents a small change, with $E$ being
small
in comparison to the total energy of the black hole.
The
change in the total
entropy of the universe is at least
\begin{equation}
   \Delta S_{total} \ge \Delta S_{BH} - S.
\end{equation}
But using the first law of thermodynamics for the black hole, this is just
\begin{equation}
\Delta S_{total} \ge \frac{E}{T_{BH}} - S = \frac{F}{T_{BH}},
\end{equation}
where $F$ is the free energy of the object at the Hawking temperature
$T_{BH}$.
In particular, since $T_{BH}>0$, the sign of $\Delta S_{total}$ must
match that of $F$. 
One concludes that the absorption of an object by a black hole can
violate the second law
only if $F<0$, in which case any of the mechanisms from section
\ref{Bb} may come into play
to prevent the process from occurring.  Note that only the first law
(energy conservation)
has been assumed in our argument and that no special properties of black
holes have been used;
the argument would proceed as well if one replaced the black hole by any
object at the same 
temperature.
(However, in places we did assume that emission and absorption
rates could be analyzed as if the black hole were in equilibrium with
its surroundings, unlike in the more general treatment of \cite{chb,tpe}.)

\subsection{The holographic bound}
\label{hologObs}

Let us now consider the holographic bound (\ref{holog}).  Suppose in
particular
that we have a (spherical, uncharged)
object with $S \ge A/4$
and consider a Schwarzschild
black hole of equal
area $A = 4 \pi R_{BH}^2$.  Since our highly
entropic object is not itself a black hole, its energy $E$
must be less than the mass $M_{BH}$ of the black hole.
The arguments of \cite{tH,LS} now ask us to consider what happens if we
drop our highly
entropic object into a black hole of mass $M_{BH}$ or otherwise
transform 
it 
into a black hole of this mass.
For arguments which drop the object into a
pre-existing black hole, one typically\footnote{See, e.g., the
weakly gravitating case described in \cite{Bek2000}.}
assumes $E \ll M_{BH}$,
but this is not the case for all the arguments.

Based on effects like those described in \cite{WaldRev}
(section \ref{SER})
and in \cite{MS} one may
speculate that some Hawking-like process
forbids this transformation.  More specifically the suggestion is that
if the transformation does proceed at
first, then the 
resulting
black hole state will be a mere `thermal fluctuation' that lasts for
no more than a time of order $R_{BH}$.
In order to 
assess this suggestion,
let us suppose for the moment that $E \ll M_{BH}$
so that the emission rate of such `highly entropic objects' from a black
hole of mass $M_{BH}$
can be analyzed as in Sections \ref{Bb} and B.
Then the free energy of our highly entropic object at
the Hawking temperature
$T_{BH} = (4 \pi R_{BH})^{-1}$ of
the black hole is
\begin{equation}
\label{hF}
F = E - T_{BH} S  < E - \frac{A/4}{4 \pi R_{BH}} = E  - M_{BH}/2 < 0.
\end{equation}
Thus, we again see that our object is
likely to be
emitted readily in Hawking radiation.

In the case where $E$ and $M_{BH}$ are comparable, the emission of our
object will react back significantly on the black hole itself.  In this
case it no longer seems possible to analyze the emission rate by
comparison with a state of thermal equilibrium in a fixed black hole
background.  (Indeed a canonical ensemble at fixed temperature seems
inappropriate, and one would have to replace it by a microcanonical
ensemble for the system of radiation plus black hole(s).)  Therefore, we
will fall back on a more general, but less compelling type of argument
which we could also have used above, but did not since the equilibrium
alternative was available.

Instead of reasoning from detailed balance and equilibrium abundances,
we could have just assumed that our object was emitted as if it were a
field quantum of a massive free field (as in the original calculations
of Hawking radiation).  This yields an emission rate, which, if we
ignore the pre-factor, takes the Boltzmannian form, $\exp(-E/T_{BH})$.
This can also be written as $\exp(\Delta S_{BH})$, where
$\Delta{S_{BH}}$ (a negative number) is the entropy lost by the black
hole in emitting the object of energy $E$, and we are still assuming
that $E\ll M_{BH}$.  If we now assume further that this rate applies
equally to each {\it microstate} of our object, then the total emission
rate for the macrostate of the object acquires a factor of $\exp(S)$,
whence the overall rate (still neglecting ``pre-factors'') takes on the
``naive thermodynamic value'' of $\exp(\Delta S_{BH} + S)$.  This
coincides with the form utilized above, $\exp(-E/T_{BH} + S)$.

Now this form of the argument has the weakness that the assmuptions
going into it seem to be under poorer control and less convincing than
those going into our equilibrium analysis above.  However, unlike the
latter, the present analysis carries over to the case where $E$ is
comparable to $M_{BH}$, at least in the sense that, according to
references \cite{kpw,mp}, 
the emission rate for a microstate retains a
factor proportional to $\exp(\Delta S_{BH})$.  If we accept this, then
the rest of the argument is just as before: The formation of a black
hole from our object could violate the second law only if $S>S_{BH}$.
But since $\Delta S_{BH}>-S_{BH}$, one finds
$\Delta{S}_{total}=\Delta{S}_{BH}+S>0$ for the corresponding emission
process.  Since this implies an exponentially large emission rate for
our object, we conclude that the combined process of collapse and
emission would actually result in a net {\it increase} in the total
entropy.

\section{Discussion}
\label{disc}

The proposals (\ref{Bek}) and (\ref{holog}) for fundamental entropy bounds
would forbid the existence of objects with extremely high entropy.
In fact, we have seen that the entropy of the putatively forbidden
objects is so high that they
(or even more entropic objects)
would be 
an important
component of Hawking radiation even for large black holes where the
temperature
is low.  This expands an interesting loophole in existing arguments for
such fundamental bounds.  In a more special context where such a
violation arises from a large number of light fields,
a very similar loophole was discussed
in \cite{WaldRev}.  To quote from that reference,
``\dots the bound should be necessary in order for black holes to be
stable or metastable states, but should not be needed for the validity
of the GSL [generalized second law].''
We found evidence for such an assertion in two different regimes.  In
the first, where the energy $E$ of the putative highly entropic object
(HEO) is much less than that of the black hole, one knows on general
grounds that the GSL cannot be violated in any semiclassical process
with a quasistationary black hole \cite{chb,tpe}.
Therefore, if one imagines a HEO which would lead to a violation, then
the conclusion must be either that the HEO cannot actually exist
(compare how self-accelerating boxes were excluded in \cite{MS})
or that some effect has been overlooked which would avoid the
violation in another way;
we presented evidence that emission of HEO's by the black hole is
such an effect. 
In the second regime, where $E$ is comparable to $M_{BH}$, things are
much  less clear cut, but a similar argument can be made if one accepts
the conclusions of \cite{kpw,mp}
concerning the emission of such objects.

One might think that such highly entropic objects are in any case
experimentally excluded due to
our excellent understanding of high temperature
thermal states produced in the laboratory.  However, states
under experimental control are produced
by interactions with normal matter.  As a result, they place only loose
constraints 
on highly entropic objects made from unknown fundamental fields which might
interact extremely
weakly with those of the standard model.
One may imagine such objects as being
made from exotic dark matter or other `hidden-sector' fields.
One might also
imagine that, even if made of standard model fields,
some dynamical effect might cause these objects to come into equilibrium
only after a cosmologically long timescale.

Since objects violating (\ref{Bek}) and (\ref{holog}) can be abundant in
Hawking radiation, it is interesting to speculate
that the production of highly entropic objects could lead to observable
rates
of mass loss from known black holes.   For example, let us consider
the case where such objects pass unimpeded through the thermal atmosphere of
the black hole but where semi-classical black holes nevertheless exist
as metastable states.  A good model for this case is the scenario of weakly
interacting Fermions discussed above.  Then if the putative bounds are
violated by a factor of order 1, our objects have negative free energy even
when their kinetic energies are relativistic.  The Hawking radiation may
then
be modeled as a `fluid' of such objects\footnote%
{Similar fluid pictures were described in \cite{CGLP} as candidate
descriptions of Hawking radiation at temperatures high enough to create
hadrons.}
which flows outward from the black hole
with density $\rho$ and speed $v \sim c$.  The black hole loses mass at a
rate of $\dot{M} = 4 \pi \rho R^2_{BH} c$.  On the other hand, we have
observed various black holes for some time and thus have at least rough
bounds on the rate at which they lose mass.  Consideration of a black hole
of a few solar masses whose mass remains roughly constant over a period
of ten years would 
rule out the existence of such a fluid with $\rho \gtrsim 5 \times 10^4
kg/m^3$, while
similar observations of a $10^6$ solar mass black hole would exclude a
corresponding 
fluid with $\rho \gtrsim 0.2 kg/m^3$.
One of course obtains much stronger limits if the accepted age of such
objects is used as the relevant timescale.
The detailed modeling of similar scenarios
may provide fertile ground for future investigations.

We conclude with a hand waving argument that also allows one to set
observational limits on certain highly entropic objects.  An enthusiastic
seminar speaker can probably wave his or her hand with an acceleration
exceeding $10^4 cm/s^2$.  If massive objects were present in the thermal
radiation associated with this acceleration then, unless these objects
were transparent to human hands, one would bump into them and the
vacuum would not feel empty.  Consider, for example, an object of mass
$\sim 1$ gram and size $\sim 1$ cm.
In order that such an object not impede our
waving hand, its 
entropy cannot exceed $10^{54}$.
This is tighter than the holographic bound by about ten orders of
magnitude, though much looser than the Bekenstein bound.
It is even possible that one can
extend this hand waving argument to rule out certain objects at zero
temperature, but we leave that discussion for another place.

\begin{acknowledgments}
D.M. would like to thank Ted Jacobson and Omer Blaes for useful discussions.
D.M. was supported in part by NSF grants PHY00-98747,  PHY99-07949,  and
PHY03-42749,
and by funds from Syracuse University, and the University of
California.
He would also like to thank the Aspen Center for Physics for their
hospitality during part of this work.
R.D.S. was partly supported by NSF grant PHY-0098488 and by a grant from the
Office of Research and Computing of Syracuse University.
\end{acknowledgments}

\end{document}